%% file: teglon.tex
\documentclass[twocolumn]{aastex701} % ,linenumbers
\usepackage{microtype}
\usepackage{amsmath}
\usepackage{booktabs}
\usepackage{tcolorbox}

\newtcolorbox{cmdbox}{
  colback=gray!5,
  colframe=gray!50,
  boxrule=0.3pt,
  arc=1mm,
  left=1mm,
  right=1mm,
  top=1mm,
  bottom=1mm
}

\DeclareMathOperator{\erf}{erf}

\def\teglon{{\tt Teglon}}

\def\Pgals{P_{\rm gals}}
\def\Pkg{P_{k, g}}

\def\ptDk{P_{{\rm 2D}_{k}}}

\def\pptDk{P^{'}_{{\rm 2D}_{k}}}
\def\ppptDk{P^{''}_{{\rm 2D}_{k}}}
\def\ppptD{P^{''}_{{\rm 2D}}}

\def\ppptDi{P^{''}_{{\rm 2D}_{i}}}

\def\Ck{C_{k}}

\def\Dk{D_{k}}
\def\Di{D_{i}}

\def\Dkg{D_{k,g}}
\def\mkg{m_{k,g}}
\def\Ltildkg{\tilde{L}_{k,g}}
\def\Wkg{W_{k, g}}

\shorttitle{\teglon: GW Follow-up Pipeline}
\shortauthors{Coulter et al.}

\graphicspath{{./}{figures/}}

\begin{document}

\submitjournal{PASP}

\title{\teglon: A Pixel-Level Pipeline for Galaxy-Informed Gravitational-Wave Follow-up Planning and Efficiency Analysis}

\author[0000-0003-4263-2228]{D.~A.~Coulter}
\affiliation{William H. Miller III Department of Physics \& Astronomy, Johns Hopkins University, 3400 N Charles St, Baltimore, MD 21218, USA}
\affiliation{Space Telescope Science Institute, Baltimore, MD 21218, USA}
\email[show]{dcoulter@stsci.edu}

\author[0000-0001-8833-474X]{P.~Darc}
\affiliation{Artificial Intelligence for Physics Laboratory (Lab-IA) and Centro Brasileiro de Pesquisas F\'{\i}sicas (CBPF), Rua Dr. Xavier Sigaud 150, Rio de Janeiro, RJ 22290-180, Brazil}
\affiliation{Center for Interdisciplinary Exploration and Research in Astrophysics (CIERA), Northwestern University, Evanston, IL 60201, USA}
\email{phelipedarc@gmail.com}

\author[0000-0002-5740-7747]{C.~D.~Kilpatrick}
\affiliation{Center for Interdisciplinary Exploration and Research in Astrophysics (CIERA), Northwestern University, Evanston, IL 60201, USA}
\email{ckilpatrick@northwestern.edu}

\author[0000-0001-7821-7195]{R.~J.~Foley}
\affiliation{Department of Astronomy and Astrophysics, University of California, Santa Cruz, CA 95064, USA}
\email{foley@ucsc.edu}

\author[0000-0002-1052-6749]{P.~McGill}
\affiliation{Space Science Institute, Lawrence Livermore National Laboratory, 7000 East Avenue, Livermore, CA 94550, USA}
\email{mcgill5@llnl.gov}

\author[0000-0003-2732-4956]{S.~D.~Wyatt}
\affiliation{Astroparticle Physics Laboratory, NASA Goddard Space Flight Center, Greenbelt, MD 20771, USA}
\email{samuel.d.wyatt@nasa.gov}

\author[0000-0002-8543-761X]{R.~Hausen}
\affiliation{Data Science and AI Institute, Johns Hopkins University, Baltimore, MD 21218, USA}
\email{rhausen@jhu.edu}

\correspondingauthor{D.~A.~Coulter}

\begin{abstract}
We describe \teglon, an open-source database and analysis pipeline engineered to optimize the search for electromagnetic (EM) counterparts to gravitational wave (GW) sources. \teglon~constructs a 3D galaxy completeness metric from an input galaxy catalog and convolves it with the 3D localization volume of a GW event, producing an updated posterior map informed by known galaxy distributions. Using \teglon, users can ingest arbitrary instrument footprints --- or download them directly from the {\tt Treasure Map} --- to generate dynamic observation plans that seamlessly interpolate between targeted galaxy pointing and region tiling. By focusing efforts on high-probability volumes, \teglon~significantly reduces the predicted search area to find EM counterparts and increases observational efficiency, especially for small field-of-view ($\lesssim1$~deg$^2$) instruments like the {\it Nancy Grace Roman Space Telescope}. Furthermore, \teglon~integrates with {\tt Redback} to calculate pixel-level model detection efficiencies, enabling custom, event-specific observing strategies or retrospective detection efficiencies on an arbitrary grid of EM transient models. \teglon~also supports custom science cases, such as prioritizing Active Galactic Nuclei searches for EM counterparts to binary black hole mergers. Having supported Gravity Collective and related programs from the LIGO-Virgo-KAGRA third observing run (O3) through O4, \teglon~is being continuously developed to support O5 and beyond. We document the public repository lineage and packaging, and provide Version~2.0 as an open-source tool for the community.
\end{abstract}


\section{Introduction}\label{sec:intro}

Ground-based laser interferometers now routinely detect gravitational waves (GW) from merging compact binaries and distribute low-latency sky maps for electromagnetic (EM) follow-up \citep[e.g.,][]{Abbott16:gw, Abbott17}. Binary neutron star (BNS) and neutron star--black hole (NSBH) mergers are predicted to produce both GW and EM radiation; the resulting optical and near-infrared counterpart, known as a kilonova, is powered by the radioactive decay of heavy elements synthesized via rapid neutron capture ($r$-process) in the merger ejecta \citep{Lattimer_Schramm74,Li98,Metzger19}. These predictions were confirmed by GW170817 \citep{Abbott17:detection}, which remains the only GW event to date with a confirmed EM counterpart: the kilonova AT\,2017gfo \citep[][also see \citealp{Abbott17}, and references therein]{Coulter17}. However, the conditions that enabled the discovery of AT\,2017gfo were exceptional, as the GW170817 source was localized within a sky region of $28\,\mathrm{deg}^{2}$ (90\% probability) at a luminosity distance of $40^{+8}_{-14}$\,Mpc. Subsequent events have proven far more challenging. For example, the second robust GW detection of a BNS, GW190425 \citep{Abbott20:gw190425}, was discovered at a greater distance ($159^{+69}_{-72}$\,Mpc) with a final localization spanning $9,881\,\mathrm{deg}^{2}$; despite a global search, no credible EM counterpart was identified by any team \citep{Coughlin19, Hosseinzadeh19, Lundquist19, Antier20, Gompertz2020, Becerra21, Chang21, Paek23, Smartt24, Coulter2025}. The scarcity of GW170817-like ``golden'' events has been reinforced by the recently completed fourth observing run (O4) of the LIGO--Virgo--KAGRA (LVK) Collaboration \citep{Acernese17:AdvVirgo,Abbott:AdvLIGO}, with the only significant NS-bearing events during O4 have been GW230518 ($\sim$460~deg$^{2}$) and GW230529 ($\sim$24,500~deg$^{2}$. Even as events are detected with increased signal-to-noise, because of detector duty cycles, localization areas can still be tens of thousands of square degrees. For example, GW230529 \citep[62\% NSBH probability;][]{Abac2024} carried a 90\% localization area of $\sim24{,}500\,\mathrm{deg}^{2}$, spanning nearly the entire accessible night sky. Full tiling of such regions is infeasible for any single facility \citep[even for Rubin, see][]{Andreoni2024}, and uninformed wide-field coverage distributes telescope time inefficiently across regions that are unlikely to host the merger.

Galaxy catalogs provide a natural astrophysical prior on where a luminous counterpart is most likely to appear \citep{Gehrels16, GTD, Adhikari_2020,2020MNRAS.492.4768D_Ducoin,Artale2020}. Since BNS and NSBH mergers are expected to trace the stellar-mass distribution of the local Universe, follow-up observations can be made more efficient by incorporating this information into the target selection process. Rather than uniformly tiling the raw LVK sky localization, observers can optimize observing schedules by convolving the 3D gravitational-wave posterior with a luminosity-weighted galaxy catalog \citep{Gehrels16,Arcavi17}.

Several community tools currently address aspects of this challenge. The {\tt Treasure Map} \citep{Wyatt20} collects, stores, and distributes user-supplied observation plans, enabling coordination across teams and reducing redundant coverage. Target and Observation Managers (TOMs; \citealp{TOM_toolkit,Coulter23,Coughlin2023_skyportal}) constitute open-source frameworks for target curation and data organization, and algorithms such as \texttt{gwemopt} \citep{Coughlin2018} or mixed-integer linear programming schedulers \citep{Singer2025} can optimize observation schedules across a network of facilities. Wide-field search campaigns during O3 and O4 frequently tiled large fractions of BNS localizations ($\geq 10^{3}\,\mathrm{deg}^{2}$) with minimal galaxy weighting \citep{Coughlin19,Coughlin20b}, relying on large field-of-view (FOV) instruments to cover these vast regions. However, in both O3 and O4, the median BNS detection range was $\lesssim 200$~Mpc \citep{GWTC3, Shah2024}, a volume within which current galaxy catalogs are already $\sim 40-50$\% complete. As a search-planning tool, \teglon\ occupies a unique, complementary niche: for wide-area localizations within this BNS horizon, the pipeline leverages this completeness information to reduce the effective search area by reweighting regions containing known galaxies within the 3D GW volume. This spatial concentration enables networks of narrow-field instruments ($\lesssim 1\,\mathrm{deg}^2$) to effectively participate in follow-up campaigns that would otherwise be observationally intractable.

Furthermore, these resulting catalog-informed HEALPix maps are designed as upstream products that TOMs, schedulers, and coordination platforms can natively consume.

We present a full public release of \teglon, a pixel-level GW follow-up optimization, planning, and analysis pipeline. First, the pipeline estimates spatially varying catalog completeness on a three-dimensional HEALPix grid. Second, it reallocates two-dimensional sky probability within each pixel based on that completeness and a per-galaxy $B$-band luminosity weighting. This produces a three-dimensional posterior over sky position and luminosity distance, augmented by per-galaxy weights (the ``4D'' map).\footnote{Throughout, we refer to the \teglon\ processed product as the ``4D'' map: the three physical dimensions of the localization posterior (sky position and luminosity distance), augmented by the per-galaxy probability weights into which the catalog-attributed fraction of each pixel's probability is redistributed. This distinguishes it from the native ``2D'' LVK sky probability.} Third, it supports model-dependent detection efficiency and upper-limit calculations using the same processed map and ingested observation footprints. \teglon\ provides a unified framework for observing programs that require catalog-informed scheduling and map-level upper limits derived from the same posterior, rather than treating sky tiling and galaxy prioritization as decoupled processes. Table~\ref{tab:usecases} summarizes the principal use cases.

In practice, the pipeline operates in two distinct modes depending on the observing program's latency requirements. In ``low-latency'' mode, \teglon\ processes maps at a fixed resolution ($\texttt{NSIDE} = 256$) and restricts analysis to the 90th percentile localization pixels to enable rapid, real-time observation planning. In ``analysis'' mode, full maps are processed at their native resolution offline to support comprehensive calculations of detection efficiencies and physical upper limits. The pipeline automatically ingests GW sky maps and metadata distributed through GraceDB or the Gravitational-Wave Transient Catalog \citep[GWTC;][]{GWTC5}, executes the reweighting procedure, and generates optimized tile plans for telescopes registered on the \texttt{Treasure Map} \citep{Wyatt20}. Real-time Target-of-Opportunity (ToO) triggers yield a fully reweighted map within 2--4\,minutes of an alert. Galaxy reweighting is most effective within the volume where the galaxy catalog is highly complete (typically $\lesssim 300\,\mathrm{Mpc}$); at larger distances, the processed map dynamically converges to the native LVK prior by design (Section~\ref{sec:real_mock_events}).

The effectiveness of this approach was demonstrated for GW190425: \teglon\ reduced the effective 90\% credible area under the host prior by concentrating the posterior into a smaller area, to $6,688\,\mathrm{deg}^{2}$, a ${\sim}1.5\times$ improvement over the raw LVK sky map, and substantially reduced the number of pointings required for small ($\leq1\,\mathrm{deg}^{2}$) FOV instruments to reach a fixed cumulative probability \citep{Coulter2025}. Similar strategies were used for the NSBH event GW190814 to rank pointings for multiple facilities \citep{Kilpatrick21}. Recently, \teglon\ has been extended to support the search for binary black hole (BBH) mergers in active galactic nuclei (AGN) environments, providing the framework for multi-model constraints on events such as S231206cc and S240413p \citep{Darc2025, darc26}.

\begin{deluxetable*}{llll}
\tablecaption{Designed \teglon\ use cases\label{tab:usecases}}
\tablehead{\colhead{Use case} & \colhead{Inputs} & \colhead{Outputs} & \colhead{Latency}}
\startdata
ToO planning & GraceDB ID / sky map & Reweighted map; tiles & 2--4\,min \\
Galaxy search & Processed map; FOV & Ranked hosts/tiles & Same night \\
Efficiency & Obs.\ manifest; models & Efficiencies; $P_M$ & Post-campaign \\
Iterative follow-up & Nondetections; depths & Viable models; plan & Multi-night \\
BBH/AGN ranking & AGN catalog + LVK map & AGN map; ranked AGN & As for ToO \\
\enddata
\tablecomments{Treasure Map FOV footprints can be registered as detectors for tiling. Analysis mode additionally requires an observation manifest with limiting magnitudes.}
\end{deluxetable*}

This paper is organized as follows. Section~\ref{sec:methodology} presents the formalism for the completeness calculation, galaxy reweighting, and model detection efficiencies at pixel resolution. Section~\ref{sec:architecture} describes the pipeline architecture and data flow. Section~\ref{sec:implementation} covers the technical implementation of \teglon\, and its usage. Section~\ref{sec:applications} summarizes past observing campaigns. Section~\ref{sec:real_mock_events} tests \teglon\ on real and mock GW events. Section~\ref{sec:results} summarizes the conditions under which \teglon\ is most useful. Section~\ref{sec:bbh} covers AGN catalog weighting for BBH follow-up. Section~\ref{sec:conclusion} concludes with repository guidance, limitations, and future work. Appendix~\ref{sec:teglon02} provides a Version~2.0 user guide.

\section{Methodology}\label{sec:methodology}
\input{teglon_appendix}

The formalism in this section was developed for earlier \teglon\ campaigns and is collected here in one place. Galaxy luminosity and distance ranking follow \citet{Coulter17}. The spatially varying completeness metric and the pixel-level detection-efficiency calculation were used for the GW190814 and GW190425 analyses of \citet{Kilpatrick21} and \citet{Coulter2025}. AGN-disk counterpart models and host ranking are those of \citet{Darc2025} and are summarized in Section~\ref{sec:analysis}. In this section, we provide the unified methodology of \teglon\ across all science cases.

\section{\teglon: Software Architecture and Data Flow}\label{sec:architecture}

\begin{figure*}
    \centering
    \includegraphics[width=\linewidth]{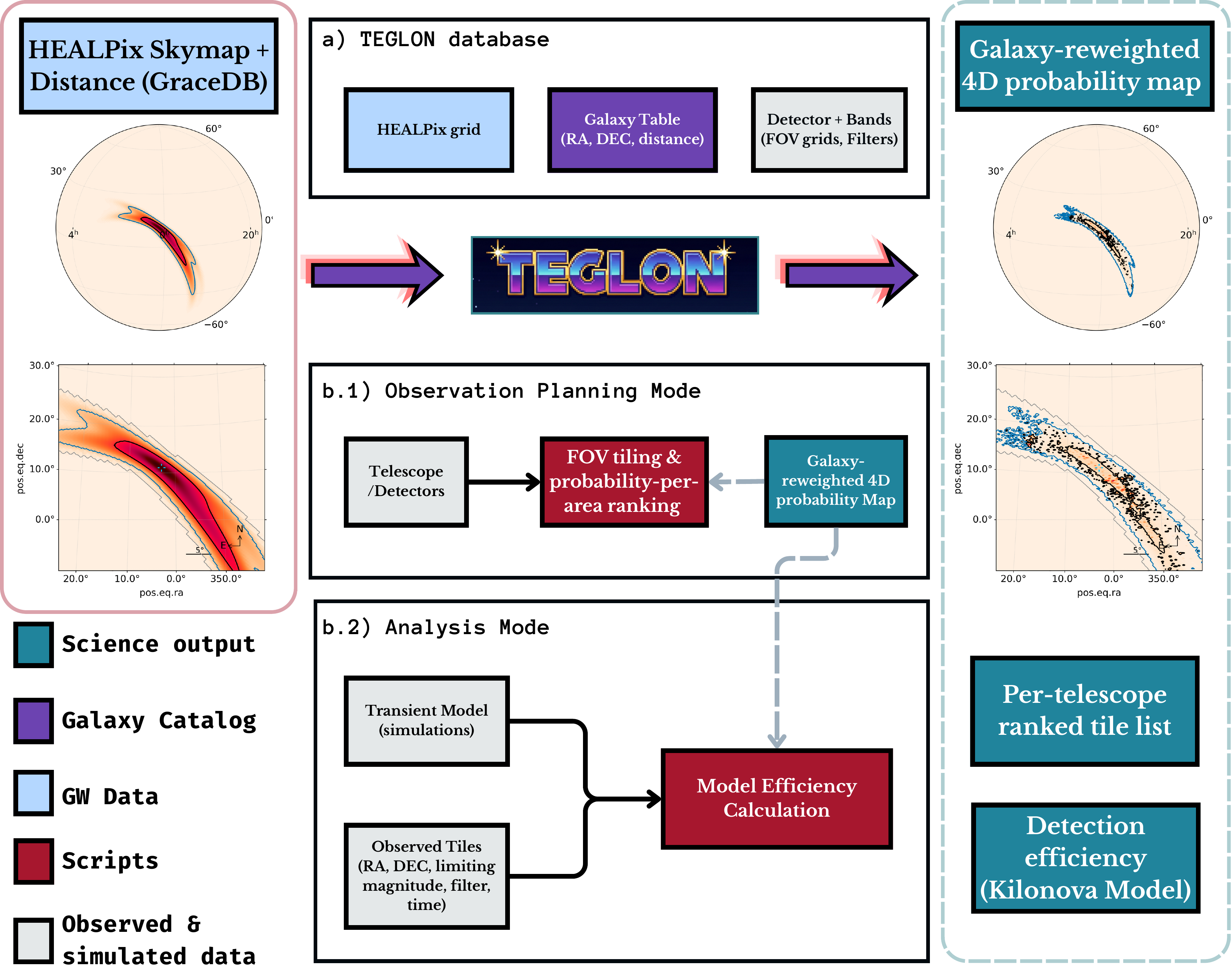}
    \caption{Schematic data flow for \teglon: LVK sky maps and catalog data are ingested into a spatial database; completeness and probability redistribution modules produce a processed HEALPix map and galaxy weights; tiling and analysis modules consume observation manifests to plan pointings and compute detection efficiencies. (a)~Database ingest of LVK sky maps and galaxy catalogs producing a galaxy-reweighted 4D probability map; (b.1)~observation-planning mode with FOV tiling and ranked tile lists; (b.2)~analysis mode combining observed tiles with transient model grids to compute detection efficiencies. The Version~2.0 CLI and container workflow are documented in Appendix~\ref{sec:teglon02}.}
    \label{fig:dataflow_v2}
\end{figure*}

At a high level, \teglon\ transforms three inputs---(i)~a GW alert and its associated HEALPix sky map, (ii)~a galaxy catalog ingested into a queryable spatial database, and (iii)~user-supplied telescope FOV geometries---into ranked galaxy lists, repixelized HEALPix maps, and telescope tiling of the GW region used for follow-up \citep{teglon_github}. The pipeline is organized around a MySQL-backed spatial database that persists across events: static science tables (galaxy catalog, Galactic dust extinction, voxel completeness, detector footprints, and the HEALPix tiling grid) are built once per catalog release, after which only event-specific tables are regenerated for each new GW alert. Figure~\ref{fig:dataflow_v2} provides a schematic overview of the \teglon\ workflow, illustrating the principal processing stages described below, the interactions between the pipeline components, and the resulting science products.

\subsection{Data Ingestion}\label{sec:ingestion}

GW localization products are distributed as multi-order HEALPix (MOC) or fixed-\texttt{NSIDE} sky maps with per-pixel distance information \citep{Singer16, GTD}. Sky maps are read with all four default fields \citep[2D probability, {\tt PROB}, and the ansatz distance parameters, {\tt DISTMU}, {\tt DISTSIGMA}, {\tt DISTNORM};][]{GTDSupplement}, and if read in low-latency mode, all values are resampled to {\tt NSIDE}~$256$. In operations, alerts and posterior products are retrieved through LVK services such as GraceDB\footnote{\url{https://gracedb.ligo.org/}} \citep{GraceDB}. For events predating the GraceDB superevent infrastructure, \teglon\ automatically queries the GWOSC Event API to recover the event epoch; the user supplies only the localization map file, and no manual GPS time entry is required.

\teglon\ ingests these maps alongside a galaxy catalog (GLADE or a successor catalog) into MySQL, where galaxies are matched to HEALPix voxels and to the static sky-pixel grid. Catalog completeness and host ranking depend on the catalog in use, so a deployment populates the static science tables once per catalog release; swapping in an updated catalog requires rebuilding voxel completeness before any science comparison is made (Section~\ref{sec:completeness_calc}). The ingestion and spatial-query logic provide the database access layer, pixel-geometry helpers, and the parsers for LVK \texttt{fits} maps and catalog-specific column mappings.

The spatial join is not evaluated geometrically at query time. At ingest, each galaxy is assigned its enclosing HEALPix pixel index at a ladder of resolutions (\textsc{nside}~$=2$--$2048$), stored as indexed integer columns, and the galaxy point geometry additionally carries a MySQL \texttt{SPATIAL} (R-tree) index; distance and redshift columns are likewise indexed. The galaxy--voxel relation is therefore materialized once as an integer-key mapping, so per-event retrievals reduce to indexed integer joins rather than repeated spherical-geometry scans, which is what allows the galaxy catalog ($\sim 1.6 \times 10^{6}$ galaxies) to be re-queried for every alert without reloading it into memory.

\subsection{Probability Redistribution}
\label{sec:redistribution}

Following the formalism detailed in Sections~\ref{sec:completeness_calc} and~\ref{sec:weighting}, each native sky-map pixel is assigned a catalog completeness factor $C_k$ derived from the voxelized galaxy catalog (Section~\ref{sec:completeness_calc}). The fraction $C_k P^{\rm 2D}_k$ of each pixel's probability is treated as attributable to cataloged galaxies and is redistributed onto per-galaxy weights using luminosity and distance priors; the remainder, $(1-C_k)P^{\rm 2D}_k$, is retained on the unstructured sky component (implemented directly as the stored quantity $\texttt{renorm2dprob}=P^{\rm 2D}_k\,(1-C_k)$). This split ensures that sight lines with low catalog completeness do not have their probability spuriously concentrated onto a sparse galaxy sample.

Probability is conserved at two distinct levels. Within a pixel the split is exact by construction: the galaxy-attributed and residual-sky terms sum identically to $P^{\rm 2D}_k$, so no renormalization is needed for the redistribution itself. Across resolutions, conservation is enforced \emph{numerically}: when a map is degraded for storage, the averaged pixel values are explicitly rescaled by the ratio of input-to-output pixel counts so that the summed probability is preserved rather than left to the averaging operator. The output is a processed HEALPix map whose pixels sum to unity (Equation~\ref{eqn:conservation}) together with a ranked list of per-galaxy probabilities, both exportable at the native \textsc{nside} or at a deliberately degraded resolution for visualization.

\subsection{Field Tiling}
\label{sec:tiling}

Tiling operates over a precomputed, per-detector grid of candidate fields (\texttt{StaticTile}), each pre-associated at build time with the HEALPix pixels it encloses (\texttt{StaticTile\_HealpixPixel}); the FOV footprints themselves are spherical polygons handled by the \texttt{Tile.py}, \texttt{Teglon\_Shape.py}, and \texttt{SQL\_Polygon.py} classes. For a given event, every static tile is assigned the summed enclosed probability of the processed map---either the ``4D'' \texttt{NetPixelProb} weight or the original pixel probability (``2D'')---and tiles are ranked in descending order of enclosed probability. A running cumulative sum (a SQL-based window function) then selects tiles until a user-specified cumulative-probability budget (constrained to $0.2$--$0.95$) or a maximum tile count is reached, subject to each detector's declination limits (\texttt{MinDec}, \texttt{MaxDec}), an optional Galactic-extinction limit, and an optional sky-region box.

The implementation prioritizes greedy coverage of the highest-probability regions, subject to user-defined overlap constraints and a maximum tile count. In practice, this approach closely matches meter-class follow-up campaigns, where schedules are iteratively updated as weather conditions, visibility constraints, and observing priorities evolve. Multiple observing plans for different facilities can be queued, and the tiling stage is intentionally decoupled from the galaxy-ranking procedure, allowing the same localization map to be used by heterogeneous facilities with different FOV footprints, observing strategies, or cadence requirements. In Version~2.0, \teglon\ can generate observing tiles for any telescope registered on the Treasure Map, enabling rapid deployment across a broad range of follow-up facilities without requiring telescope-specific modifications to the underlying workflow.

\subsection{Priority Ranking}
\label{sec:ranking}

Tiles and galaxies are ordered by their integrated posterior weight. The constraints applied at this stage are those represented in the database: per-detector declination limits (\texttt{MinDec}/\texttt{MaxDec}), a Galactic-extinction limit derived from the SFD dust map (the per-tile $E(B-V)$, capped by a user \texttt{-{}-extinct} limit), the photometric band, and an optional rectangular sky region. Ranked outputs can be filtered by cumulative enclosed probability---retaining, for example, the smallest set of galaxies whose summed weights reach 50\% or 90\% of $P_{\rm gals}$---and are written as per-telescope ranked tile lists, while the underlying ranked galaxy weights persist in the database for query and export. This ranking step enabled Gravity Collective programs to preferentially observe high-$B$-luminosity hosts inside the GW localization of GW190814 while still covering most of the 90\% credible region \citep{Kilpatrick21}.

\subsection{Analysis Mode and Simulations}
\label{sec:analysis}

Beyond real-time planning, \teglon\ supports an analysis mode. Once a list of completed exposures is ingested---each as an \texttt{ObservedTile} record specifying a field's central coordinates and footprint polygon, filter (\texttt{Band}), mid-time (\texttt{MJD}), exposure time, and limiting magnitude (\texttt{Mag\_Lim})---the pipeline recomputes, pixel by pixel, the detection efficiency of a chosen transient model against the achieved depths, and hence yields model-dependent upper limits, using the same processed map (Section~\ref{sec:upper_limits}). \teglon\ implements kilonova, GRB-afterglow, and linear light-curve models (\texttt{-{}-model\_type kne|grb|linear}); for each pixel it compares the model's apparent magnitude at the event epoch---propagated through the pixel's marginalized distance and corrected for Milky Way extinction---against the limiting magnitude actually reached there. The calculation determines, given only the exposures that were actually taken, which kilonova or afterglow simulations would have been detectable, and therefore which regions of parameter space can be excluded without acquiring any new data. Synthetic depths or cadences can be explored by editing the observation list. The polygon--pixel intersection used here reuses the same \texttt{healpy}-based machinery as tiling, ensuring internal consistency between the planning and analysis stages.

In Version~2.0, \teglon\ optionally supports any transient model available in the \texttt{Redback} package \citep{REDBACKSarin2024}. Given the name of a \texttt{Redback}-implemented photometric model---such as \texttt{one\_component\_kilonova} or \texttt{two\_component\_kilonova}---the extension draws $N_\mathrm{sim}$ samples from the default \texttt{Redback} priors and writes each as a \teglon-format ECSV \texttt{.dat} file for analysis mode. Users may optionally modify the priors, number of simulations, and time interval; a usage guide is provided in the online documentation.\footnote{\url{https://github.com/gw-commons/teglon/tree/main/redback_extension}}

The built-in one-component kilonova model integrated into \teglon\ follows the semi-analytic prescription of \citet{Villar17}, parameterized by the ejecta mass $M_{\mathrm{ej}}$, ejecta velocity $v_{\mathrm{ej}}$, and gray opacity $\kappa$. Two opacity values are simulated, $\kappa = 0.5\,\mathrm{cm^{2}\,g^{-1}}$ and $\kappa = 10\,\mathrm{cm^{2}\,g^{-1}}$ (hereafter the ``blue'' and ``red'' components), generated as two distinct sets of simulations with the photospheric floor temperature held fixed at 6000\,K and 2500\,K, respectively. GRB-afterglow models are generated with the \texttt{afterglowpy} package \citep{Ryan20}, adopting a structured jet observed on-axis and off-axis. The jet microphysics and geometry are held fixed across the simulations with $\eta_0 = 8.02$, $\Gamma_B = 12.0$, $\epsilon_e = 0.1$, $\epsilon_B = 6.761\times10^{-6}$, $p = 2.15$, and $\xi_N = 1.0$, while the isotropic-equivalent kinetic energy $E_{K,\mathrm{iso}}$, circumburst density $n$, and observer viewing angle $\theta_{\mathrm{obs}}$ are varied on a grid: $E_{K,\mathrm{iso}} \in [0.01, 1000]$, $n \in [10^{-6}, 10]\,\mathrm{cm^{-3}}$, and $\theta_{\mathrm{obs}} \in \{0^\circ, 17^\circ\}$ (on-axis and off-axis). The jet's energy is a fixed fraction of the isotropic-equivalent value, $E_0 = 4.33965\times10^{-4}\,E_{K,\mathrm{iso}}$, held constant across the grid.

More recently, we expanded the scope of \teglon\ to work on simulated EM counterparts to BBH mergers. The majority of gravitational-wave detections reported by the LVK Collaboration correspond to BBH mergers, which are not expected to produce EM emission in vacuum. However, several theoretical studies have shown that observable EM counterparts may arise if the merger occurs within a dense gaseous environment, such as the accretion disk of an active galactic nucleus (AGN) \citep{Bartos17, Antoni19, Grobner20, Kaaz2023}.

We have implemented three such scenarios in \teglon, all based on the general idea that asymmetric BBH mergers (in mass and/or spin) receive a recoil velocity (or kick) due to anisotropic gravitational-wave emission. As the kicked remnant travels through the AGN disk at supersonic speeds relative to the surrounding gas, it can undergo super-Eddington Bondi--Hoyle--Littleton accretion, powering radiation-driven outflows and/or relativistic jets. The first scenario \citep{McKernan19} models the thermal emission from an off-center hotspot produced when the remnant black hole carries its gravitationally bound gas through the AGN disk. Ram pressure strips most of this gas, shock-heating it and producing a transient thermal signal. The second scenario \citep{Tagawa24} considers a relativistic jet launched as the merger remnant accretes disk gas; we simulate the thermal emission during jet breakout, as well as the subsequent cooling of both the disk and the jet cocoons. The third scenario \citep{RodriguezRamirez24} considers a BBH merger occurring outside the AGN thin disk; the kicked remnant subsequently intersects the disk at an angle $\theta_k$ relative to the disk normal, and we simulate the thermal emission from material driven out of the disk by the resulting jet-inflated cocoons. A detailed description of each model's parameters and the priors used to generate the simulated grids, and of their application to S231206cc, can be found in \citet{Darc2025} (see also Section~\ref{sec:bbh}). The BBH simulations from the models of \cite{McKernan19,Tagawa24,juan_publicado,juan_novo}, required for the analysis mode, are publicly available through the repository.

\section{Technical Implementation}\label{sec:implementation}

The public codebase is written primarily in Python~3 (3.9--3.11) and is installable as a package (\texttt{teglon-o4}) that exposes a single \texttt{teglon} command-line entry point; runtime dependencies are declared in \texttt{pyproject.toml} and pinned to the container build in \texttt{docker/teglon\_worker/requirements.txt}, with optional legacy plotting dependencies isolated in a \texttt{[legacy]} extra. A \texttt{Dockerfile} and \texttt{docker/docker-compose.yml}, wrapped by a \texttt{./teglon} helper script, provide a containerized deployment that bundles a MySQL service, so no database port-forwarding is required for the pipeline\footnote{See \href{https://github.com/gw-commons/teglon/tree/main/docs}{\teglon\ Documentation}}. Core scientific libraries include \texttt{healpy} for pixel operations \citep{Zonca19, Gorski05}, \texttt{astropy} where used for coordinates and units \citep{astropy}, and MySQL for persistent storage of galaxies and observation metadata. Spatial joins (galaxy--voxel, polygon--pixel) are expressed as SQL queries over spatially indexed sky-position and distance columns so that repeated re-tile operations during a night do not reload the full catalog into memory.

Instrument footprint polygons (registered as detectors), filter throughputs, and per-epoch limiting magnitudes supplied by the user in an observation manifest drive the detection-efficiency calculations in Section~\ref{sec:upper_limits}; polygon--pixel intersection reuses the same HEALPix machinery as tiling.

\subsection{Usage and Command-Line Interface}

Routine operations run through the unified \texttt{teglon} command-line interface, which exposes each stage of the workflow as a subcommand. A typical run (1)~registers a GW event identifier and produces the reweighted (``4D'') localization using \texttt{teglon trigger}. For public GW superevents, the user only needs to provide the event identifier; \texttt{teglon} automatically queries \texttt{GraceDB}, retrieves the corresponding sky map, performs the galaxy reweighting procedure, and outputs an updated HEALPix probability map in FITS format, typically in 2--4 minutes per event; (2)~exports ranked per-telescope tile lists with \texttt{teglon extract}, while the underlying ranked galaxy weights persist in the database; and (3)~generates FITS and PDF/PNG sky maps with \texttt{teglon plot} and \texttt{teglon compare}. Auxiliary commands support routine operations: \texttt{teglon setup} performs the one-time database build, \texttt{teglon doctor} runs dependency and configuration checks, and a \texttt{-{}-json} mode provides machine-readable output for scripting, follow-up scheduling, and observing-planning workflows. Operators typically maintain one long-lived database per catalog release and only refresh event-specific sky map tables when a new LVK localization arrives (or remove an event with \texttt{teglon delete-event}). Beyond the CLI, all stages remain importable Python modules---the \teglon\ driver class (\texttt{teglon.py}) together with the core objects \texttt{Completeness\_Objects.py}, \texttt{Pixel\_Element.py}, and \texttt{Detector.py}---allowing observatories to embed \teglon\ within broker systems or dynamic scheduling services, similar in spirit to other transient-management platforms \citep{Coulter23}. For teams that only need exposure-level bookkeeping and visualization against the LVK map, complementary tools remain valuable \citep{Wyatt20}; the niche of \teglon\ is providing end-to-end consistency between catalog-informed localizations and observational-efficiency accounting.

\section{\teglon~Motivation and Past Applications}\label{sec:applications}

The \teglon\ algorithm was derived from the rapid-response techniques used for the Swope Telescope discovery of the optical counterpart of GW170817 \citep{Coulter17} -- specifically the method of weighting individual galaxies, and of packing multiple galaxies into single FOVs -- and has since supported a broad network of telescopes employed in the search for EM counterparts to GW events. Below we highlight representative programs where at least one of the authors of this paper participated; in each case, galaxy-weighted prioritization and/or map-level efficiency accounting of the type now packaged in \teglon\ was used. Table~\ref{tab:pastevents} summarizes \teglon-relevant products for the events discussed here.

\begin{deluxetable*}{llcccl}
\tablecaption{\teglon-relevant products for events discussed in this work\label{tab:pastevents}}
\tablehead{\colhead{Event} & \colhead{Class} & \colhead{Native 90\%} & \colhead{\teglon\ product} & \colhead{Mode} & \colhead{Ref.} \\
 & & \colhead{($\mathrm{deg}^{2}$)} & & & }
\startdata
GW170817 & BNS & 28 & Precursor pipeline & Plan & \citet{Coulter17} \\
GW190425 & BNS & 9881 & Eff.\ 90\% cred.\ area: 6688\,$\mathrm{deg}^{2}$ & Plan+Analysis & \citet{Coulter2025} \\
GW190814 & NSBH & 37.7 & Ranked 1M2H hosts & Plan+Analysis & \citet{Kilpatrick21} \\
S231206cc & BBH & 445.1 & AGN-disk constraints & AGN+Analysis & \citet{Darc2025} \\
S240413p & BBH & 38.2 & AGN opt./spec.\ follow-up & AGN+Plan & \citet{darc26} \\
\enddata
\tablecomments{Native 90\% areas are LVK credible regions measured from the maps used in this work: GW170817 from the preferred LALInference map \citep{Abbott17:detection}; GW190425 from the final GWTC-2 map \citep{Coulter2025} (distinct from the BAYESTAR map in Table~\ref{tab:gw_summary_compact}); GW190814, S231206cc, and S240413p match the one-decimal native areas in Table~\ref{tab:gw_summary_compact}. ``Eff.\ 90\% cred.\ area'' denotes the effective 90\% credible area under the host prior after \teglon\ reweighting. Full pointing-level \teglon\ archives for every facility are beyond the scope of this software release.}
\end{deluxetable*}

\subsection{GW170817 and the 1M2H}

The first binary neutron star merger with an electromagnetic counterpart was discovered in an $i$-band image taken by the 1~m Swope telescope at Las Campanas Observatory \citep{Coulter17}. This discovery was enabled by a combination of galaxy targeting and galaxy clustering (since ported to \teglon), and resulted in the rapid localization of the counterpart in the 9th search image $\lesssim12$~hours after GW merger. This fast discovery also enabled the first spectrum \citep{Shappee17} and subsequent multi-color light-curve analysis by \citet{Drout17, Kilpatrick17}.

\subsection{GW190425 (BNS Candidate)} \label{sec:GW190425}
The candidate neutron star binary GW190425 was localized to $9881\,\mathrm{deg}^{2}$ at 90\% confidence in the native LVK sky map, motivating a galaxy-targeted search with the Gravity Collective. Applying \teglon\ galaxy reweighting reduced the effective 90\% credible area under the host prior to $6688\,\mathrm{deg}^{2}$ \citep{Coulter2025}. \citet{Coulter2025} present a comprehensive analysis of the imaging campaign---including limiting magnitudes, cadence, and kilonova priors---with \teglon\ completeness weighting applied to GLADE hosts. Facilities contributing to that effort included the Lick 0.76-m Katzman Automatic Imaging Telescope (KAIT), the Las Cumbres Observatory Global Telescope (LCOGT) 1~m telescope network, the 1-m Swope telescope, the 1-m Nickel telescope, the 0.7-m Thacher telescope, and the 10-m Keck~I telescope, which performed near-infrared imaging \citep{Kilpatrick21, Coulter2025, mosfire:instr}.

\subsection{GW190814 (NSBH)}
For the asymmetric-mass compact binary event GW190814, localized to $37.7\,\mathrm{deg}^{2}$ at 90\% credibility in the final LVK sky map \citep{Abbott20} as measured for Table~\ref{tab:gw_summary_compact}, Gravity Collective observations again combined KAIT, LCOGT, Swope, the Nickel telescope, Thacher, and Keck facilities to cover the localization and characterize candidate transients \citep{Kilpatrick21}. \teglon\ was used explicitly to rank galaxies inside the LVK map for 1M2H pointings and to place model-dependent limits from the nondetection \citep{Kilpatrick21}.

\subsection{Binary Black Hole Mergers in AGN Environments}

Long-baseline optical and spectroscopic monitoring of BBH candidates spatially coincident with bright AGN hosts provides a unique opportunity to test models of disk-embedded BBH mergers. By using \teglon in its analysis mode, \citet{Darc2025} presented multi-model constraints on possible electromagnetic emission from GW231206 a BBH candidate with a Bilby 90\% credible localization area of $445.1\,\mathrm{deg}^{2}$, while \citet{darc26} reported optical imaging and spectroscopic follow-up of GW240413\_022019 ($38.2\,\mathrm{deg}^{2}$ at 90\% credibility), enabling constraints on the expected detectable emission based on parameters inferred from spectroscopic data. Beyond motivating the AGN catalog extensions described in Section~\ref{sec:bbh}, these studies established a methodology for model-driven follow-up of BBH mergers by constraining the regions of parameter space most likely to produce detectable electromagnetic counterparts under different AGN-driven BBH scenarios.

Rather than relying solely on rapid, deep observations, these analyses demonstrate that BBH follow-up benefits from combining astrophysically motivated simulations with long-baseline monitoring to probe the delayed and diverse emission predicted by current models. More broadly, they illustrate how \teglon\ can be used not only to optimize observational planning but also to interpret follow-up data in the context of physically motivated BBH electromagnetic emission models.

\begin{table*}[t]
\centering
\caption{Summary of \teglon\ Performance and Pointing Reductions Across Selected Events}
\label{tab:gw_summary_compact}
\setlength{\tabcolsep}{4pt}
\begin{tabular}{l c c c c}
\toprule
\textbf{Event} & \textbf{GW190425} & \textbf{GW190814} & \textbf{S240413p} & \textbf{S231206cc}$^{\ast}$ \\
\midrule
$D_L$ [Mpc] & $154.62 \pm 44.62$ & $275.59 \pm 55.82$ & $509.84 \pm 114.71$ & $1260.52 \pm 293.86$ \\
$\langle C \rangle$ & 0.454 & 0.843 & 0.170 & 0.004 \\
\midrule
\multicolumn{5}{l}{\textbf{Localization Area Reduction [$\mathrm{deg}^{2}$]}} \\
50\% Area & $2805.7 \rightarrow 1016.4\ (-63.8\%)$ & $7.5 \rightarrow 3.0\ (-59.2\%)$ & $12.8 \rightarrow 11.8\ (-7.4\%)$ & $117.9 \rightarrow 117.7\ (-0.2\%)$ \\
90\% Area & $10182.6 \rightarrow 8697.3\ (-14.6\%)$ & $37.7 \rightarrow 24.0\ (-36.4\%)$ & $38.2 \rightarrow 38.0\ (-0.7\%)$ & $445.1 \rightarrow 444.9\ (-0.0\%)$ \\
\midrule
\multicolumn{5}{l}{\textbf{Pointings (50\% / 90\% credible)}} \\
Roman & $8672\rightarrow5293$ / $31838\rightarrow28820$ & $24\rightarrow24$ / $121\rightarrow107$ & $39\rightarrow38$ / $122\rightarrow122$ & $364\rightarrow363$ / $1422\rightarrow1421$ \\
Rubin/LSST & $306\rightarrow257$ / $1110\rightarrow1078$ & $2\rightarrow3$ / $6\rightarrow8$ & $3\rightarrow3$ / $7\rightarrow8$ & $15\rightarrow15$ / $57\rightarrow57$ \\
T80-S & $3044\rightarrow3013$ / -- & $5\rightarrow6$ / $22\rightarrow22$ & $20\rightarrow\text{--}$ / -- & $95\rightarrow94$ / -- \\
\bottomrule
\end{tabular}
\tablecomments{Performance comparison of native LVK and \teglon-reweighted localizations across four campaigns.
$D_L$ is the luminosity distance (mean $\pm$ $1\sigma$ uncertainty).
$\langle C \rangle$ is the average host-galaxy catalog completeness across the 3D localization volume.
Arrows ($X \rightarrow Y$) denote the transition from the native LVK parameter ($X$) to the catalog-informed \teglon\ value ($Y$), with percentage reductions shown in parentheses.
Pointing counts are evaluated at both the $50\%$ and $90\%$ credible regions for the \textit{Nancy Grace Roman Space Telescope} ({\it Roman}; $\sim 0.28\,\mathrm{deg}^2$ FOV), the Vera C. Rubin Observatory (Rubin/LSST; $\sim 9.6\,\mathrm{deg}^2$ FOV), and the T80-South telescope (T80-S; $\sim 2\,\mathrm{deg}^2$ FOV).
For GW190425, the table evaluates the initial low-latency \texttt{BAYESTAR} sky map circulated during the live trigger to document real-time planning performance; this is distinct from final offline analyses utilizing the GWTC-2 posterior.
Dashes (---) indicate configurations that were not computed.  For S231206cc, we note that the mean distance to the event of $\approx$1260~Mpc is beyond the nominal 1200~Mpc threshold for {\tt Teglon}'s galaxy catalogs.  {\tt Teglon} still operates in this regime for the fraction of the event's localization that is within 1200~Mpc, and we assume zero catalog completeness outside of this distance threshold, demonstrating the flexibility of our algorithm for events near this threshold.}
\end{table*}

\section{Real and Simulated Gravitational-Wave Events}\label{sec:real_mock_events}

In this section we evaluate \teglon\ through three complementary experiments, each designed to isolate and study a single variable. First, we apply \teglon\ to the four real events discussed throughout this work, at their true distances and localization geometries, quantifying the posterior concentration achieved under operational conditions. Second, we fix all four localization geometries at a common, favorable distance of 100\,Mpc and vary only the telescope FOV, isolating how the benefit depends on instrument footprint. Third, we hold the two-dimensional geometry fixed and vary the assumed source distance from 40 to 1000\,Mpc, isolating the dependence on catalog completeness. Our main objective is to quantify how strongly the galaxy-informed redistribution concentrates the localization posterior, measured as the reduction in credible area and in the number of telescope pointings needed to reach a fixed cumulative probability, and how that concentration mainly depends on the follow-up facilities employed and the source distance.

Unless noted otherwise, ``LVK'' and ``\teglon'' curves and pointing counts use the same precomputed static-tile grid, declination limits, and cumulative-probability ranking for each facility; only the enclosed pixel weights differ (native 2D probability versus the completeness-reweighted \texttt{NetPixelProb} map). To probe the impact of telescope FOV, we adopt three representative facilities spanning wide, intermediate, and narrow footprints: the Vera~C.~Rubin Observatory Legacy Survey of Space and Time (Rubin/LSST; \citealt{2019ApJ...873..111I}), with a $9.62\,\mathrm{deg}^{2}$ FOV; the T80-S telescope used by S-PLUS, with a $2\,\mathrm{deg}^{2}$ FOV \citep{t80_Davis2023,t80_Santos2024,T80_JacobsonGaln2024,t80_teglon_Hu2025,Darc2025,T80_Schroeder2025,T80_Rastinejad2025,T80_Teixeira2026,darc26}; and the {\it Nancy Grace Roman Space Telescope} \citep{ROMAN2021AAS...23732701M}, with a $0.281\,\mathrm{deg}^{2}$ FOV. These span ground- and space-based observatories; the Rubin/LSST and Roman plans are demonstration tilings rather than claims of executed ToO campaigns.

\subsection{Real Events}\label{sec:results_real}

Table~\ref{tab:gw_summary_compact} summarizes, for the four real GW events, the reduction in credible area and the corresponding catalog completeness $\langle C\rangle$, together with the pointings each facility requires to enclose the 50\% and 90\% credible regions. The events span a wide range of distance and completeness. Distances quoted here and in Table~\ref{tab:gw_summary_compact} are the map-derived luminosity-distance means used for the tiling experiment.

For the nearby BNS merger GW190425 ($D_L \simeq 155\,\mathrm{Mpc}$; $\langle C\rangle = 0.45$), the 50\% credible area contracts from 2806 to $1016\,\mathrm{deg}^{2}$ ($-64\%$), reducing the Roman requirement from 8672 to 5293 pointings and the Rubin/LSST requirement from 306 to 257.\footnote{The GW190425 entries in Table~\ref{tab:gw_summary_compact} reweight the initial BAYESTAR localization circulated during the live search (90\% area $10182.6\,\mathrm{deg}^{2}$, $D_L = 154.62\pm44.62\,\mathrm{Mpc}$; \citealt{Singer19_GCN24168}), so as to document performance on the product available in real time. This is distinct from the $9881\rightarrow6688\,\mathrm{deg}^{2}$ figures quoted in the abstract, Table~\ref{tab:pastevents}, and Section~\ref{sec:GW190425}, which use the final GWTC-2 map \citep{Coulter2025}; \teglon\ reweighting is applied identically, and the difference reflects only the input localization.} For the more distant NSBH event GW190814 ($D_L \simeq 276\,\mathrm{Mpc}$; $\langle C\rangle = 0.84$), the reweighting contracts the 50\% and 90\% credible areas by 59\% and 36\%, respectively, yet the pointing counts change only marginally because the localization is already compact. For the S240413p BBH candidate ($D_L \simeq 510\,\mathrm{Mpc}$; $\langle C\rangle = 0.17$), the 50\% credible region requires 38 Roman pointings against 39 for the native map, while the Rubin/LSST requirement is unchanged at three. Finally, for the distant BBH candidate S231206cc ($D_L \simeq 1.26\,\mathrm{Gpc}$; $\langle C\rangle = 0.004$), the reweighted and native posteriors yield essentially identical areas and pointing counts. The pattern across the four events---large gains for the nearby, wide localization and negligible change for the distant, low-completeness ones---is examined systematically in Section~\ref{sec:results_distance}.

\begin{figure*}
    \centering
    \includegraphics[width=1\linewidth]{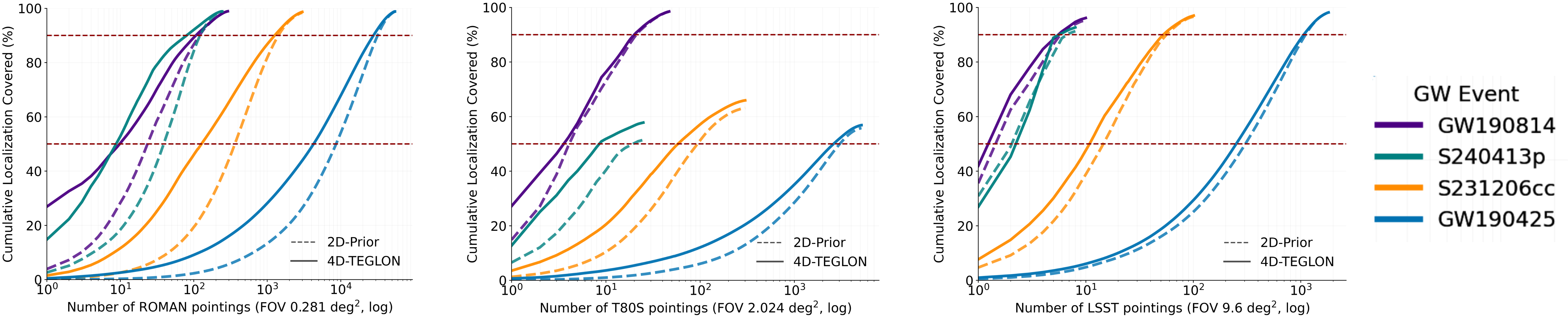}
    \caption{Cumulative localization probability as a function of the number of telescope pointings for three representative facilities: the {\it Nancy Grace Roman Space Telescope} (\emph{left}), T80-S (\emph{center}), and Rubin/LSST (\emph{right}). Each curve ranks the same static tiles by one map and accumulates that map’s own probability: native LVK (2D; solid) or \teglon\ (4D; dashed). The plot therefore shows how many pointings are needed to enclose a given fraction of each posterior. Red dashed lines mark the 50th and 90th percentile credible levels. The events are mock realizations at 100,Mpc.}
    \label{fig:coverage_comparison}
\end{figure*}

\subsection{Field-of-View Dependence at Fixed Distance}\label{sec:results_fov}

To further investigate the dependence of the method on the telescope FOV, we generated mock GW events but fixed the two-dimensional localization geometry of each real event and placed it at a common luminosity distance of 100\,Mpc, with a 20\% distance uncertainty. This favorable, fixed distance holds catalog completeness high for all four geometries, so that any variation in the measured gain reflects the telescope FOV for different localization geometries rather than the source distance. The four geometries---those of GW190425, GW190814, S231206cc, and S240413p---were chosen because they span the range of localization morphologies and sky areas commonly encountered in GW follow-up campaigns.

Figure~\ref{fig:coverage_comparison} shows the cumulative localization probability enclosed as a function of the number of telescope pointings for each facility. As expected, the benefit of \teglon\ grows for instruments with smaller FOVs, where observing efficiency is driven by the ability to prioritize the highest-probability regions within the localization volume. For S240413p, the Roman Space Telescope requires only $\sim9$ pointings to cover the 50\% credible region under the \teglon\ posterior, a 77\% reduction relative to the native LVK localization; T80-S required 55\% fewer pointings. The gain is modest for wide-field facilities such as Rubin/LSST: for GW190425, Rubin/LSST---with its $9.62\,\mathrm{deg}^{2}$ FOV---requires 306 pointings on the native map and 252 under the \teglon\ posterior, reflecting the smaller relative improvement afforded by its large sky coverage. (The corresponding real-distance GW190425 values in Table~\ref{tab:gw_summary_compact}, $306\rightarrow257$, differ only because that experiment adopts the true $\sim155\,\mathrm{Mpc}$ distance rather than the fixed $100\,\mathrm{Mpc}$ used here.)

\begin{figure*}[t]
    \centering
    \includegraphics[width=0.8\linewidth]{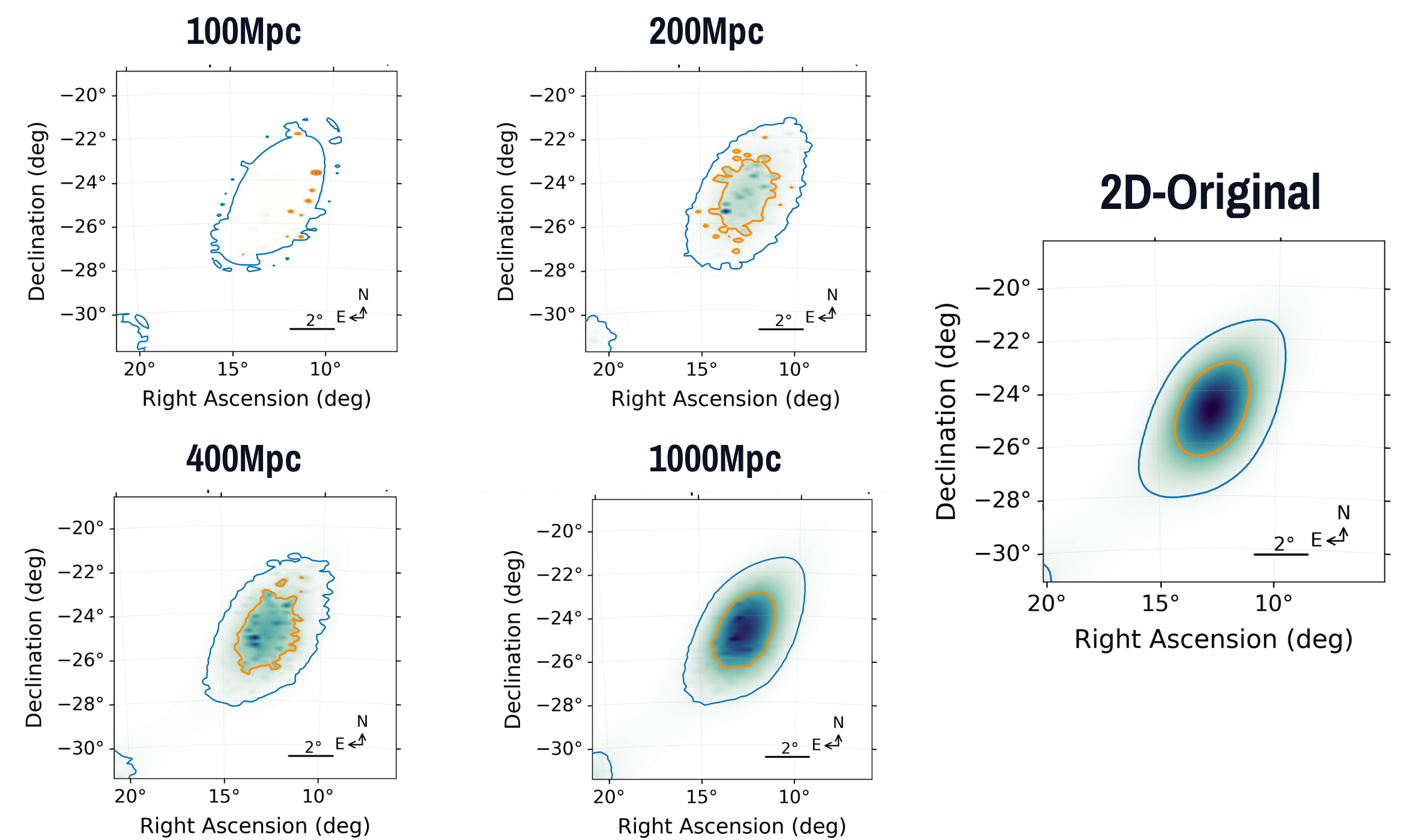}
    \caption{Galaxy-informed redistributions for the GW190814 sky geometry, holding the two-dimensional localization fixed while shifting the assumed source distance. The right panel is the native LVK (``2D original'') map; the others are \teglon\ reweighted maps at 100, 200, 400, and 1000\,Mpc. Orange and blue contours mark the 50\% and 90\% credible regions, respectively. The 100\,Mpc panel is recentered and zoomed on the concentrated high-probability region (different RA/Dec range), so the secondary low-probability island visible in the shared window of the other panels falls outside that frame. Each panel includes a $2^{\circ}$ scale bar. At large distance the reweighted map converges toward the native prior as catalog completeness declines.}
    \label{fig:alldist_events}
\end{figure*}

\begin{figure*}[t]
    \centering
    \includegraphics[width=0.99\linewidth]{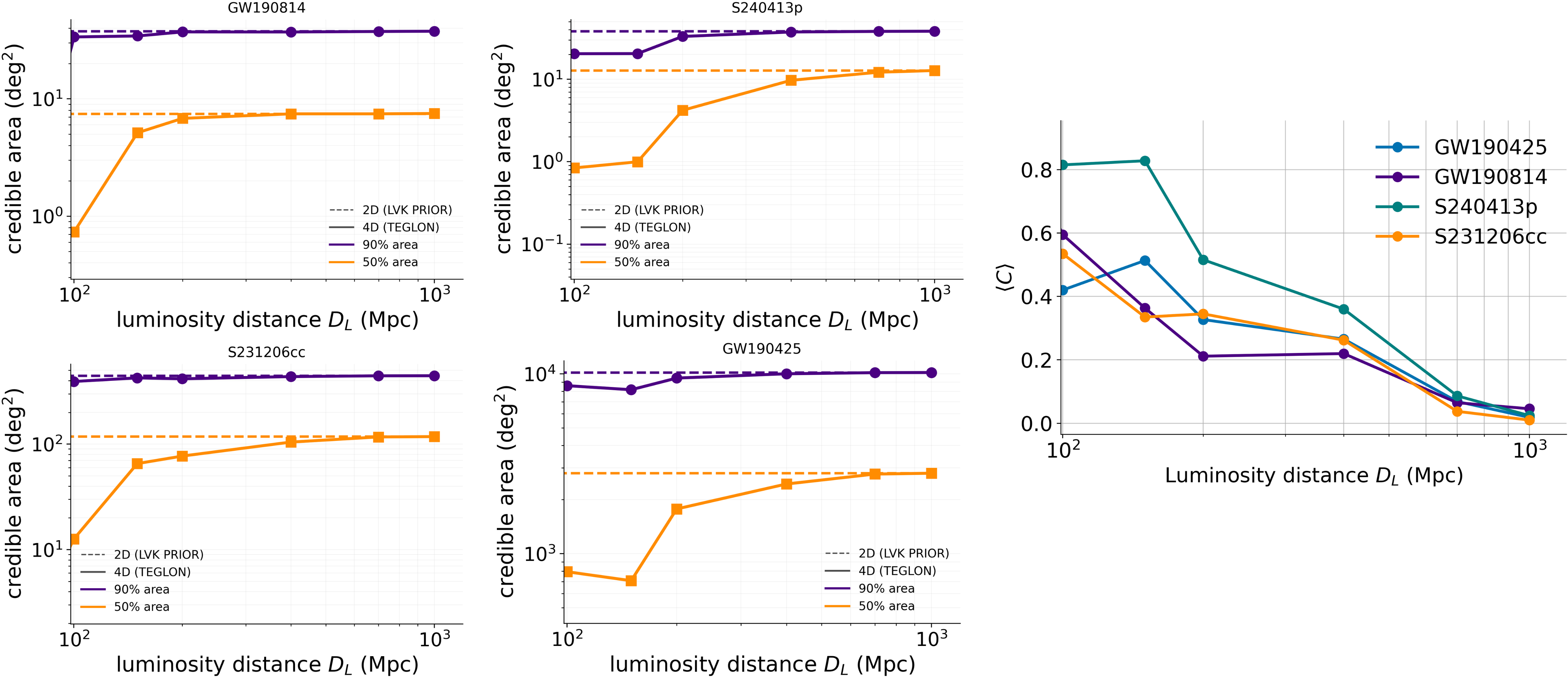}
    \caption{Evolution of the 50\% and 90\% credible areas as a function of assumed source distance for mock realizations based on GW190425, GW190814, S231206cc, and S240413p, together with the corresponding catalog completeness $\langle C\rangle$ along each localization. All maps are evaluated at a fixed HEALPix resolution of $\texttt{NSIDE}=256$ ($\approx 0.23^{\circ}$ per pixel). As in Figure~\ref{fig:alldist_events}, gains relative to the native LVK prior shrink as completeness declines.}
    \label{fig:areaallxdist}
\end{figure*}

\subsection{Distance and Completeness Dependence}\label{sec:results_distance}

The effectiveness of \teglon\ depends strongly on source distance and catalog completeness (Section~\ref{sec:completeness_calc}): the redistribution relies on the completeness of the underlying galaxy catalog within the localization volume, so nearby events benefit most, while the improvement declines as the source distance grows and completeness falls. To characterize this behavior directly and as a consistency check, we generated a second mock suite of GW events in which each two-dimensional geometry is held fixed and the assumed source distance varies from 40 to 1000\,Mpc, with a fiducial distance uncertainty of 20\%.

Figure~\ref{fig:alldist_events} illustrates the resulting redistributions at representative distances relative to the native two-dimensional LVK map for the GW190814 geometry. Figure~\ref{fig:areaallxdist} traces the 50\% and 90\% credible areas as a function of assumed distance across the full four-event suite, together with the completeness $\langle C\rangle$ along each localization. As the assumed distance increases, the reweighted localization converges toward the native GW prior, approaching it by $\sim1000,\mathrm{Mpc}$, as seen in both Figures~\ref{fig:alldist_events} and~\ref{fig:areaallxdist}. Catalog completeness declines, so a smaller fraction of the posterior is assigned to cataloged galaxies at these large distances.  The 20\% distance window also grows, averaging over more galaxies and reducing contrast among those that remain. These effects combine to cause the \teglon\-reweighted localization to converge to the original GW localization.

\subsection{Physically Informed Follow-up}
\label{sec:physically_informed}

Having quantified how reweighting concentrates the localization posterior, we next describe how the same processed map supports an iterative, model-informed observing strategy.

Beyond static tiling of a fixed localization map, \teglon\ supports an iterative follow-up procedure in which each night of imaging updates the set of viable electromagnetic counterparts and thereby reshapes the next night of observing. The required ingredients are already present in the pipeline: a completeness-reweighted HEALPix localization (Sections~\ref{sec:weighting} and~\ref{sec:redistribution}), an observation manifest of achieved depths (Section~\ref{sec:analysis}), and a grid of transient models whose detectability can be evaluated pixel by pixel (Section~\ref{sec:upper_limits}).

Consider a BNS or NSBH alert for which Night~1 yields no optical counterpart. Each exposure in the observation manifest is stored as an \texttt{ObservedTile} record with central coordinates, footprint polygon, filter (\texttt{Band}), mid-time (\texttt{MJD}), exposure time, and limiting magnitude (\texttt{Mag\_Lim}). For a model light curve $M_\lambda(t)$, \teglon\ computes the maximum detectable luminosity distance in each covered pixel,
\begin{equation}
D_{i, M, \lambda}(t)~\mathrm{(Mpc)} = 10^{0.2\, [\mu_{i, M, \lambda}(t) - 25]},
\end{equation}
where the distance modulus is $\mu_{i, M, \lambda}(t) = m_{i, \lambda}(t) - M_{\lambda}(t) - A_{i, \lambda}$, $m_{i, \lambda}(t)$ is the achieved limiting magnitude, and $A_{i, \lambda}$ is the Milky Way extinction in that band. The pipeline then integrates the pixel distance posterior from $D=0$ to $D_{i, M, \lambda}(t)$ under a Gaussian approximation with mean $\Di$ and standard deviation $\sigma_{\Di}$ (Equation~\ref{eqn:dist_integral}). Combining epochs with the complement of the product of nondetection probabilities (Equation~\ref{eqn:complement_product}) and combining filters (Equation~\ref{eqn:model_prob}) yields a campaign detection probability $P_M$ for that model given only the exposures that were actually taken.

For kilonova follow-up the default grid follows the one-component semi-analytic models of \citet{Villar17}, parameterized by ejecta mass $M_{\mathrm{ej}}$, ejecta velocity $v_{\mathrm{ej}}$, and gray opacity $\kappa$. \teglon\ provides two opacity sequences, $\kappa = 0.5\,\mathrm{cm^{2}\,g^{-1}}$ (``blue'') and $\kappa = 10\,\mathrm{cm^{2}\,g^{-1}}$ (``red''), with photospheric floor temperatures fixed at $6000\,\mathrm{K}$ and $2500\,\mathrm{K}$, respectively (Section~\ref{sec:analysis}). The same analysis mode also accepts GRB-afterglow grids generated with \texttt{afterglowpy} \citep{Ryan20} and linear light-curve options (\texttt{-{}-model\_type kne|grb|linear}). Models with $P_M$ above a chosen exclusion threshold, for example $P_M \geq 0.9$, would have been recovered with high probability under Night~1 conditions. Their nondetection therefore rules out that region of parameter space, conditional on the GW candidate being astrophysical and on the adopted light-curve family. The surviving models are those with $P_M$ below the threshold.

Those surviving realizations are then propagated forward in time. Because the Villar grids are tabulated as absolute magnitude versus rest-frame phase, each allowed $(M_{\mathrm{ej}},v_{\mathrm{ej}},\kappa)$ point predicts apparent magnitudes in the available filters at Night~2, Night~3, and later epochs after applying the pixel distance posterior and extinction already stored with the processed map. Observers can therefore maximize the expected detection probability $\langle P_M\rangle$ over the surviving ensemble when choosing the next night of filters, exposure times, and fields. In practice this favors fields that enclose large fractions of the reweighted localization, filters where the remaining models are brightest relative to achievable depths, and cadences matched to the predicted decline. Blue components with $\kappa = 0.5\,\mathrm{cm^{2}\,g^{-1}}$ typically evolve on $\sim$1--2\,day timescales, whereas red components with $\kappa = 10\,\mathrm{cm^{2}\,g^{-1}}$ can remain detectable for several days \citep{Villar17, Metzger19, Li98}.

This procedure is especially useful for the wide localizations characteristic of O3--O4 BNS and NSBH candidates. For GW190425, the native LVK 90\% credible area was $9881\,\mathrm{deg}^{2}$. \teglon\ reduced the effective 90\% credible area under the host prior to $6688\,\mathrm{deg}^{2}$, a factor of ${\sim}1.5$ improvement, and reduced the number of pointings required to reach a fixed cumulative probability for instruments with FOV $\leq 1\,\mathrm{deg}^{2}$ by roughly a factor of two to several, depending on footprint and distance \citep{Coulter2025}. The Gravity Collective campaign that followed, using KAIT, LCOGT, Swope, Nickel, Thacher, and Keck/MOSFIRE, placed model-dependent limits on kilonova emission with \teglon\ completeness weighting applied to GLADE hosts \citep{Kilpatrick21, Coulter2025, mosfire:instr, Dalya18}. For the NSBH event GW190814, the same ranking machinery directed 1M2H pointings while the nondetection constrained ejecta energetics \citep{Kilpatrick21}. In both cases the scientific product was a quantitative assessment of which transient models remained allowed, together with the ranked tiles. Feeding that assessment back into the planner couples nondetections to the next observing plan: Night~$N$ nondetections update $P_M$ across the model grid, and Night~$N{+}1$ tiles are then drawn from the processed map with weights informed by the models that have not yet been ruled out.

The computational cost is compatible with overnight or same-night updates. A typical map rebuild with \texttt{teglon trigger} completes in 2--4 minutes per event. Efficiency evaluation with \texttt{teglon efficiency} is parallel over pixels and model realizations (Appendix~\ref{sec:teglon02} illustrates runs with \texttt{-{}-num-cpu 70}). New limiting magnitudes can therefore be ingested between nights, or between epochs on the same night when a rapidly evolving blue kilonova is still a viable target.

Two caveats limit the quantitative interpretation. First, exclusions inherit the assumptions of the adopted model grid and of the Gaussian distance approximation in Equation~\ref{eqn:dist_integral}. They constrain the chosen family of light curves rather than all possible electromagnetic counterparts. Second, at large distance or low catalog completeness the reweighted map converges to the native LVK prior (Section~\ref{sec:real_mock_events}). In that regime the spatial component of the plan reverts toward uniform-in-probability tiling, even while the choice of filter and cadence can remain model-informed. Within those limits, nondetections become actionable constraints: they prune parameter space, forecast the photometry of what remains, and concentrate subsequent telescope time where a detection is still plausible given both the GW localization and the accumulated depth of the campaign.

\section{Discussion}\label{sec:results}

The applications presented above, together with the mock-data analysis, demonstrate that map-level completeness modeling materially changes where follow-up time is spent relative to uniform tiling, particularly for the wide LVK localizations typical of O2--O3 BNS candidates \citep{Kilpatrick21, Coulter2025}. Gains are largest for nearby events and for instruments with small FOVs, where ranking high-probability galaxies and tiles can reduce the number of pointings needed to reach a fixed cumulative probability (Section~\ref{sec:real_mock_events}).

Additionally, for NSBH systems the same machinery bounds kilonova energetics when no counterpart is found \citep{Kilpatrick21}. BBH campaigns instead use \teglon\ primarily as a scheduler and analysis framework for studies of AGN-driven BBH mergers \citep{Darc2025, darc26}. Campaign science results (counterpart searches, flux limits, and model exclusions) remain in the cited papers; this work documents the map-level products those analyses share. Users should treat every processed map as tied to both the LVK sky map version and the catalog ingest: archival comparisons between O3 and O4 campaigns are only meaningful when those inputs are held fixed or explicitly reprocessed.

\section{Specialized Catalog Additions: EM Counterparts to Binary Black Holes}\label{sec:bbh}

Before the start of the fourth observing run (O4), public-facing LVK materials quoted an illustrative forecast of $260^{+330}_{-150}$ BBH mergers per year in O4 (LVK public-alerts documentation; \citealt{Abbott:Local}). During the first segment of O4 (O4a; $\sim$7.7~months of science-quality data, not counting three BBH events from engineering time preceding O4a), the LVK Collaboration reported 75 BBH events, i.e.\ $\sim$117~yr$^{-1}$, consistent with the lower part of pre-O4 expectations. Updated population analyses using GWTC-4.0 find a local BBH merger rate of $14$--$26~\mathrm{Gpc}^{-3}~\mathrm{yr}^{-1}$ (90\% credible interval) with substantial astrophysical and catalog incompleteness caveats \citep{Abbott25:GWTC4Pop}. While BBH mergers are not generally expected to produce luminous electromagnetic transients on their own, galaxy nuclei should host the densest populations of stellar-mass black hole binaries, and some of these binaries may reside in AGN disks. Gas-rich environments can supply torques and dissipation that accelerate inspiral \citep{Bartos17, Antoni19, Grobner20, Kaaz2023} and produce mergers in baryon-rich regions where shocks or flares are more plausible. An optical flare discovered $\sim$34~days after GW190521 (S190521g) was proposed as a counterpart \citep{Graham20, Abbott20:GW190521}; see \citealt{Ashton20} for a critical discussion of the association.

To support BBH follow-up, \teglon\ can optionally ingest the AGN catalog of \citet{Secrest15}, which lists 1.4~million AGNs to $g\approx 26$~mag selected from AllWISE \citep{Wright2010}. That catalog is estimated to be $\gtrsim 84\%$ complete for known AGNs and for all AGNs with $R<19$~mag; for $z<0.1$ it is expected to be $\gtrsim 90\%$ complete. Operationally, the AGN table is loaded in place of (or alongside) the default GLADE ingest and the voxel-completeness and galaxy--pixel associations are rebuilt once before science runs (Appendix~\ref{sec:teglon02}); subsequent \texttt{teglon trigger} / \texttt{extract} commands then rank AGN hosts inside the LVK volume using the same redistribution machinery as for BNS/NSBH hosts (Sections~\ref{sec:redistribution} and~\ref{sec:analysis}). This weighting scheme specifically targets environments associated with BBH mergers in AGN disk models and underpins the S231206cc and S240413p analyses cited in Section~\ref{sec:applications}.

\section{Conclusion}\label{sec:conclusion}

\teglon\ provides a self-consistent map-level treatment of catalog completeness, galaxy-based probability concentration, and imaging detection efficiency for GW follow-up, supporting alert-night planning, retrospective efficiency accounting, and optional AGN-host weighting within a single processed HEALPix representation. The historical public codebase is documented at \citep{teglon_github}; O4-era packaging and the Version~2.0 CLI described in Appendix~\ref{sec:teglon02} (Figure~\ref{fig:dataflow_v2}) are developed in the \texttt{teglon} repository (\url{https://github.com/gw-commons/teglon/tree/main/docs}), which descends from the earlier \texttt{teglon\_O4} fork. Future work includes deeper integration with survey-specific systematics, expanded host libraries beyond GLADE, regression tests against archived LVK sky maps, and streamlined APIs for observatory brokers.

\subsection{Limitations and Future Development}

Remaining systematics include catalog depth anisotropy (Section~\ref{sec:completeness_calc}), Gaussian approximations to distance posteriors (Section~\ref{sec:upper_limits}), greedy static-tile ranking relative to a full multi-facility scheduler, and the need to refresh voxel completeness whenever GLADE, Secrest, or successor catalogs are updated. Demonstration tilings for Rubin/LSST and Roman (Section~\ref{sec:real_mock_events}) illustrate FOV dependence and do not substitute for executed survey ToO strategies. Planned development includes API documentation, automated tests against reference sky maps, configurable weighting for specialized hosts (Section~\ref{sec:bbh}), and expanded validation on next-generation survey scenarios. Source code, container recipes, and issue tracking are public; we encourage users to cite both this paper and the repository release tag or commit hash used in any publication that quotes ranked lists, sky maps, or efficiencies derived from the pipeline.

\subsection{Software and Data Availability}

\teglon\ is open source. The Version~2.0 install path used in this work is \url{https://github.com/gw-commons/teglon/tree/main/docs} (see Appendix~\ref{sec:teglon02}). Completeness figures use the GLADE catalog \citep{Dalya18}; BBH host ranking optionally uses the AllWISE AGN catalog of \citet{Secrest15}. LVK sky maps are public via GraceDB and GWOSC. Transient model grids included in the repository follow \citet{Villar17}, \citet{Ryan20}, and the BBH scenarios of \citet{McKernan19}, \citet{Tagawa24}, and \citet{RodriguezRamirez24,juan_novo} as implemented in \citet{Darc2025,darc26}.

\begin{acknowledgments}

We thank the LIGO Scientific, Virgo, and KAGRA Collaborations for open data products and low-latency sky maps. We acknowledge the builders and maintainers of GLADE, HEALPix, and the Python ecosystem libraries on which \teglon\ depends. This work benefited from discussions within the Gravity Collective and the broader GW follow-up community.

This work originated at UCSC; D.A.C. acknowledges support from the National Science Foundation (NSF) Graduate Research Fellowship.

C.D.K.\ gratefully acknowledges support from the NSF through AST-2432037, the HST Guest Observer Program through HST-SNAP-17070 and HST-GO-17706, and from JWST Archival Research through JWST-AR-6241 and JWST-AR-5441.

P.D.\ acknowledges support from the Artificial Intelligence for Physics Laboratory (Lab-IA) and the Centro Brasileiro de Pesquisas F\'{i}sicas (CBPF).

P.M. acknowledges that this work was performed under the auspices of the U.S. Department of Energy by Lawrence Livermore National Laboratory under Contract DE-AC52-07NA27344. The document number is \textcolor{red}{LLNL-JRNL-XXXXX}.

\end{acknowledgments}

\vspace{5mm}
\facilities{Las Cumbres Observatory (LCOGT), Lick Observatory:KAIT, Las Campanas Observatory:Swope, Nickel (Lick), Keck (I, II), Thacher Observatory}

\software{Astropy \citep{astropy}, HEALPix \citep{Gorski05}, healpy \citep{Zonca19}, afterglowpy \citep{Ryan20}, MySQL, \teglon\ \citep{teglon_github}}

\appendix

\section{\teglon\ Version 2.0 User Guide}\label{sec:teglon02}
Version~2.0 preserves the scientific machinery described above but repackages \teglon\ for reproducible, single-command operation. The multi-stage workflow---which previously required invoking individual scripts inside a container and tunneling to the database---is consolidated behind a single command-line entry point (\texttt{teglon}) backed by a containerized MySQL service. Database credentials are resolved from environment variables, allowing the client to run entirely within the container network and eliminating the need for host port forwarding or database tunnels. Every legacy per-stage script remains functional, leaving existing workflows unaffected. The release includes a \texttt{pyproject.toml} for a \texttt{pip}-installable client, a Docker wrapper, a unit-test suite, and updated documentation. The complete workflow was validated end-to-end on the five GW events discussed here: GW170817, GW190425, GW190814, S240413p, and S231206cc.

\paragraph{Installation and one-time database build.}
The science database (galaxy catalog, dust extinction, detector footprints, and the static tiling grid) is built only once. On a typical multi-core server this step requires approximately 45--63\,min in our tests,\footnote{Timed on a dedicated host with 96 CPU cores and large local disk; wall times scale with available cores and storage bandwidth.} after which only event-specific tables are generated. Specialized catalogs (e.g., Secrest AGN hosts; Section~\ref{sec:bbh}) are enabled by swapping the ingest ASCII dump and re-running the completeness build before science comparisons.

\begin{cmdbox}
\ttfamily\footnotesize
git clone https://github.com/gw-commons/teglon.git \\
cd teglon\_O4\_updated
cp docker/.env.example docker/.env\\
cp Settings.example.ini Settings.ini\\
./teglon up\\
./teglon setup --run
\end{cmdbox}

\paragraph{Single-command processing.}
The core operation maps a GW event identifier (e.g., S240413p) to a physically informed sky map based on the galaxy weighting scheme. The \texttt{trigger} command retrieves the localization, redistributes probability onto galaxies (Section~\ref{sec:weighting}), and writes the processed HEALPix map to disk in approximately 2--4 minutes per event.

\begin{cmdbox}
\ttfamily\footnotesize
./teglon trigger S240413p
\end{cmdbox}

This produces, for example,

\begin{cmdbox}
\ttfamily\footnotesize
web/events/S240413p/S240413p\_4D\_reweighted\_bayestar.fits.gz
\end{cmdbox}

\paragraph{Observation mode.}
Field tiling, ranking, and visualization can be executed either as individual stages or through the unified \texttt{run} pipeline. Outputs include telescope-specific ranked tile and galaxy catalogs (ECSV), together with all-sky observation plans.

\begin{cmdbox}
\ttfamily\footnotesize
./teglon run S240413p
\end{cmdbox}

Alternatively, the workflow can be executed stage-by-stage:

\begin{cmdbox}
\ttfamily\footnotesize
./teglon load-map S240413p\\
./teglon extract S240413p --cum-prob 0.9 --num-tiles 1000\\
./teglon plot S240413p
\end{cmdbox}

\paragraph{Analysis mode.}
Given a list of telescope pointing data (RA, Dec, $3\sigma$ limiting magnitude), the processed localization map can be used to compute pixel-level detection efficiencies and model-dependent upper limits (Section~\ref{sec:upper_limits}). Transient model grids are distributed under \texttt{web/models/}.

\begin{cmdbox}
\ttfamily\footnotesize
./teglon load-obs S240413p --tile-file pointings.ecsv\\
./teglon efficiency S240413p --model-type kne --num-cpu 70
\end{cmdbox}

\paragraph{Transient models.}
\teglon\ currently implements kilonova models for BNS and NSBH mergers based on the light-curve templates of \citet{Villar17}, together with GRB-afterglow and linear light-curve options (Section~\ref{sec:analysis}).

\paragraph{Catalog and instrument management.}

New facilities can be registered directly from their Treasure Map footprint using the command below.

\begin{cmdbox}
\ttfamily\footnotesize
./teglon add-telescope --tm-detector-id <ID> \\
\end{cmdbox}

\paragraph{Pre-superevent localizations.}
For events predating the GraceDB superevent infrastructure (e.g.\ GW170817), the client automatically falls back to the GWOSC Event API to recover the event epoch. The user only provides the localization map; no manual GPS time entry is required.

\begin{cmdbox}
\ttfamily\footnotesize
./teglon trigger GW170817 --healpix-file MCMC\_TF2\_LowSpin\_AllSky.fits
\end{cmdbox}

The user can also inspect any sky map using the \texttt{healpy} package \citep{Zonca19}:
\begin{cmdbox}
\ttfamily\footnotesize
./teglon skymap-info web/events/[GW\_ID]/skymap.fits
\end{cmdbox}

\paragraph{Host (non-container) deployment.}
The client is also distributed as a standard Python package. Database configuration is controlled through environment variables, allowing identical commands to run against local or remote MySQL instances without modification.

\begin{cmdbox}
\ttfamily\footnotesize
pip install -e .\\
export DATABASE\_HOST=127.0.0.1\\
export DATABASE\_PORT=53306\\
export DATABASE\_PASSWORD=<password>\\
teglon trigger S240413p
\end{cmdbox}

\paragraph{Backward compatibility.}\label{sec:teglon02-compat}
All original functionality is preserved. The new commands act as thin wrappers around the unchanged stage-level scripts, ensuring complete backward compatibility with previous workflows. For example, the one-time initialization performed by

\begin{cmdbox}
\ttfamily\footnotesize
./teglon setup --run
\end{cmdbox}

is equivalent to the sequence of legacy commands:

\begin{cmdbox}
\ttfamily\scriptsize
docker compose run --rm gw\_script python ./web/src/utilities/bulk\_upload\_glade.py

docker compose run --rm gw\_script python ./web/src/utilities/initialize\_teglon.py \\
--build\_skydistances --build\_skypixels --build\_detectors \\
--build\_TM\_detectors --build\_bands --build\_MWE \\
--build\_galaxy\_skypixel\_associations --build\_completeness \\
--compose\_completeness --build\_static\_grids

docker compose run --rm gw\_script python ./web/src/utilities/build\_init\_pickles.py \\
--build\_skypixels\_pickle --build\_ebv\_pickle \\
--build\_composed\_completeness\_pickle
\end{cmdbox}

\bibliography{teglon}
\bibliographystyle{aasjournalv7}

\end{document}

%% file: teglon_appendix.tex
\subsection{Galaxy Catalog and Completeness Calculation}
\label{sec:completeness_calc}

In the third LIGO/Virgo observing run (O3), the One-Meter Two-Hemispheres (1M2H) collaboration used the Galaxy List for the Advanced Detector Era (GLADE) \citep{Dalya18} galaxy catalog for electromagnetic search, follow-up, and analysis of GW events \citep{Kilpatrick21} because GLADE is substantially more complete than the Gravitational Wave Galaxy Catalog (GWGC) at the LVK O3 median binary neutron star (BNS) inspiral range. GLADE combines and cross-matches sources from the GWGC, the HyperLEDA catalog \citep{Makarov14}, the Two Micron All Sky Survey Extended Source Catalog \citep{Skrutskie06}, the 2MASS Photometric Redshift Catalog \citep{Bilicki13}, and the Sloan Digital Sky Survey Data Release 12 quasar catalog \citep{Paris2017}. GLADE contains $\sim 2.96 \times 10^{6}$ galaxies, $\sim 2.95 \times 10^{5}$ quasars, and $149$ globular clusters.

The spatial distribution of GLADE galaxies is notably anisotropic \citep[see Fig.~1 in][]{Dalya18} because of large-scale structure, reduced completeness along the Galactic plane, and differing number densities from each component survey/contributing database (HyperLEDA alone contributes $\sim$2.6~million sources). \teglon\ uses this anisotropic completeness to decide when to emphasize galaxy-targeted observations over uniform localization tiling via a spatially varying 3D completeness metric.

To compute this completeness metric, we use the {\tt GLADE+} catalog \citep[see][]{Dalya22} from its ASCII distribution into a MySQL database, where each original catalog column is mapped to a column in the \texttt{Galaxy} table. We then apply a series of selection criteria. First, we retain only objects classified as galaxies, removing approximately 9.1\% of the catalog entries. Next, we exclude galaxies for which either the redshift and apparent $B$-band magnitude are unavailable. After these quality cuts, the ingested catalog has 1,614,426 galaxies (49.46\% of the full GLADE catalog).

\begin{figure}
    \centering
    \includegraphics[width=0.99\linewidth]{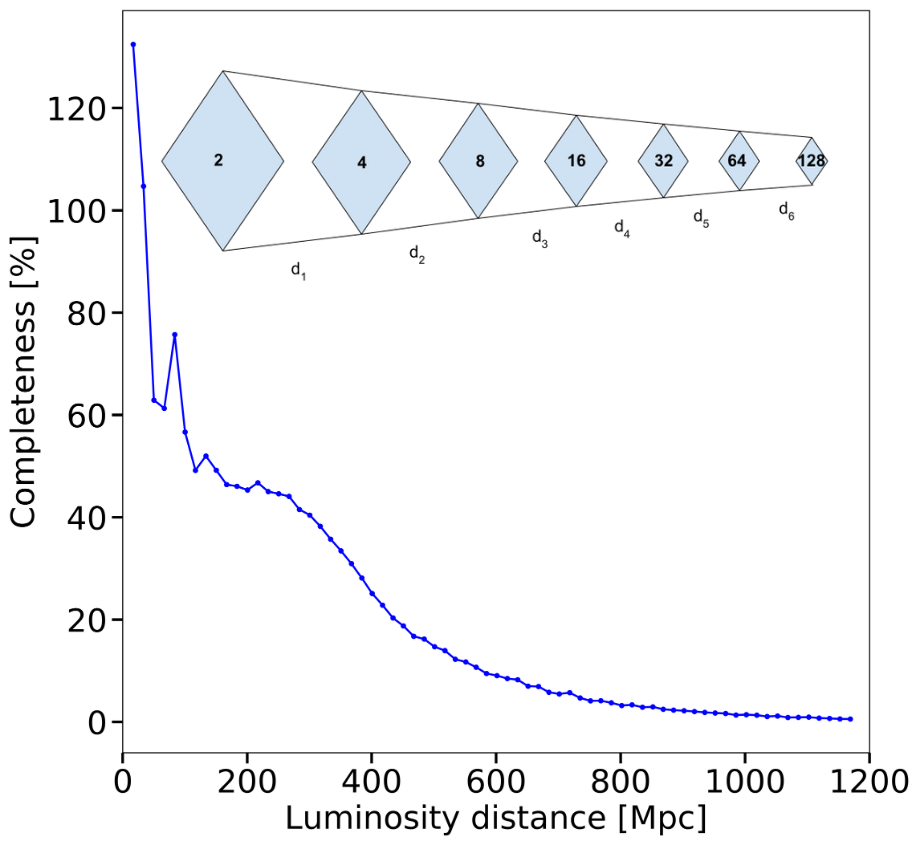}
    \caption{Simplified schematic of the resolution and distance steps through a sight line in the \teglon\ sky pixel grid (\emph{top; not to scale}) and uncorrected completeness values computed from the enclosed voxel blue luminosity along a representative sight line (\emph{bottom}).}
    \label{fig:voxels}
\end{figure}

We define an adaptive, three-dimensional all-sky grid using the Hierarchical Equal Area isoLatitude Pixelization (HEALPix) scheme\footnote{\url{https://healpix.sourceforge.io}} \citep{HEALPix}. This framework partitions the sphere into equal-area pixels and serves as the standard format for LVK sky maps \citep{Singer16, GTD}. The resulting volumetric grid is composed of prism-like cells (``voxels'') whose transverse faces are HEALPix pixels, with radial boundaries that bracket successive distance bins (see Figure~\ref{fig:voxels}). We select the HEALPix resolution and radial binning so that each voxel subtends approximately equal, average comoving volume ($\sim1.5\times10^3$~Mpc$^{3}$). The grid volume is bounded by $D \in [0,1200]$~Mpc, with the HEALPix resolution $N_{\mathrm{side}}$ scaling dynamically from 2 ($\leq45$~Mpc) to 128 ($\geq900$~Mpc; see Figure~\ref{fig:voxels}) to yield 196{,}608 independent sight lines. Every galaxy with $D \leq 1.2$~Gpc is assigned to its enclosing voxel. For each sight line at $N_{\mathrm{side}}=128$, we query Milky Way $E(B-V)$ using an SFD map via \texttt{dustmaps}\footnote{\url{https://dustmaps.readthedocs.io}} \citep{ebv, Green18} and convert to band-specific extinction following \citet{Schlafly11}.

We define a completeness scalar $C$ as the ratio of the voxel $B$-band luminosity density to the mean blue luminosity density in the local Universe \citep{Kopparapu08}. Each voxel $i$ corresponds to a sky position and luminosity distance, $C_i = C(\alpha_i, \delta_i, D_{L,i})$. Given $g$ galaxies in voxel $i$, we convert each galaxy's apparent $B$ magnitude $m_g$ to absolute $B$ luminosity using its luminosity distance $D_g$ (Mpc) and line-of-sight extinction $\lambda_g$ in the $B$ band,
\begin{align}\label{eq:mu_L}
    \mu_g &= 5\log_{10} D_g + 25 ,  \\
    L_g &= 10^{-0.4\,(m_g - \mu_g - \lambda_g)} .
\end{align}

\noindent Summing over galaxies in the voxel gives the total $B$-band luminosity. Following \citet{Kopparapu08}, we express luminosities in units of $L_{10} \equiv 10^{10}\,L_{B,\odot}$, with $L_{B,\odot} = 2.16 \times 10^{33}$~erg~s$^{-1}$ and $M_{B,\odot} = 5.48$ \citep{BandT08},
\begin{equation}
    L_{10,i} = \frac{\sum_{j} L_j}{L_{B,\odot} \times 10^{10}} .
\end{equation}

\begin{figure}[t]
    \centering
    \includegraphics[width=\columnwidth]{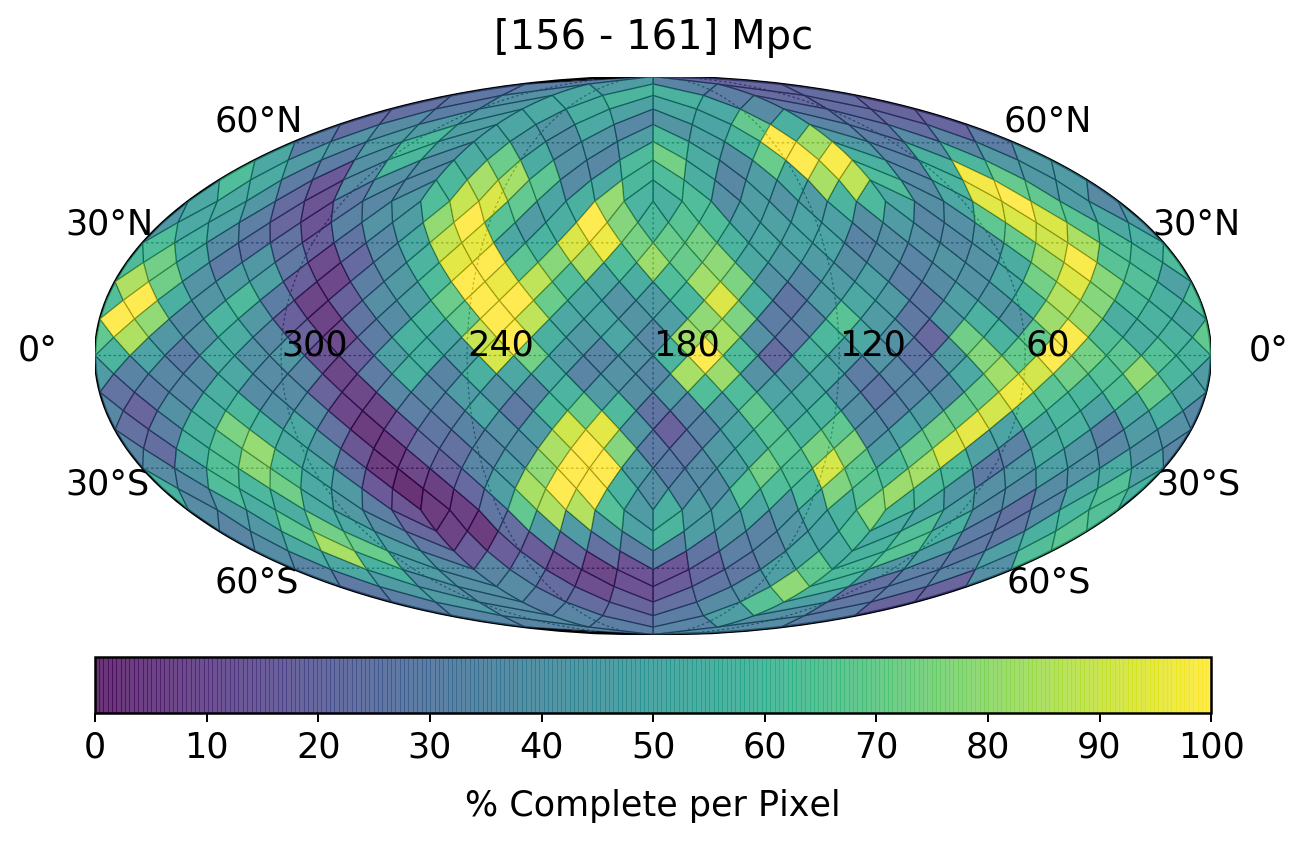}
    \caption{All-sky completeness visualization from \teglon\ for the distance bin corresponding to GW190425's luminosity distance, $\sim159$~Mpc. The horseshoe-shaped feature of systematically lower completeness corresponds to the Milky Way plane.}
    \label{fig:completeness}
\end{figure}

\noindent Dividing $L_{10,i}$ by the voxel volume $V_i$ yields the voxel $B$-band luminosity density; dividing that by $1.98 \times 10^{-2}\,L_{10}\,\mathrm{Mpc}^{-3}$ \citep{Kopparapu08} gives the uncorrected completeness for that voxel. Because the mean density is global, nearby voxels ($D \leq 40$~Mpc) can exceed 100\% owing to the local overdensities. Coarse radial binning at small $D$ can also produce discontinuities between adjacent pixels. For use as search weights we therefore cap $C$ at unity and apply symmetric Gaussian smoothing on the sphere with the \texttt{healpy}\footnote{\url{https://healpy.readthedocs.io}} \citep{Zonca19, Gorski05} library using a $30$~Mpc transverse radius at each distance slice. Figure~\ref{fig:completeness} shows the all-sky completeness at the distance bin enclosing GW190425.

\subsection{Completeness-Weighted 2D Probability Redistribution}
\label{sec:weighting}

When the LVK detects a compact-binary coalescence and produces rapid parameter estimation, the sky location and distance are represented as a posterior sampled on an adaptive HEALPix grid \citep{Singer16}. Each pixel carries a marginalized distance distribution and a two-dimensional sky probability.

Upon ingesting the native LVK sky map, \teglon\ maps the localization posterior to our custom HEALPix grid. We partition the native probability of each pixel into two components: a fraction $\Ck\,\ptDk$ is treated as associated with cataloged galaxies, while the remaining fraction $(1-\Ck)\ptDk$ is retained on the uncataloged sky. The resulting residual sky probability is $\pptDk$ is left in the original pixel while $\Ck\,\ptDk$ is collected from each pixel to make up the probability budget to be allocated for redistribution, $\Pgals$. The split in probabilities is given by:

\begin{align}
    \pptDk &= \ptDk \left( 1 - \Ck \right), \\
    \Pgals &= \sum_{k} \ptDk\, \Ck .
\end{align}

\noindent \teglon\ distributes $\Pgals$ over galaxies using the scheme of \citet{Coulter17}. Pixel $k$ contains $g$ galaxies, each with luminosity distance $\Dkg$ and apparent $B$ magnitude $\mkg$. The luminosity weight of the $g$th galaxy in the $k$th pixel is
\begin{equation}\label{eq:6}
    \Ltildkg = \Dkg^{2}\, 10^{-0.4\, \mkg}.
\end{equation}

\noindent We also compute a distance weight to boost the salience of galaxies located at the $k$th pixel's mean GW distance,

\begin{equation} \label{eqn:distance_weight}
    \tilde{D}_{k, g} = 1 - \erf\!\left( \frac{\lvert \Dkg - \Dk \rvert}{\sqrt{\sigma_{\Dkg}^{2} + \sigma_{\Dk}^{2}}} \right),
\end{equation}
with $\sigma_{\Dkg}$ and $\sigma_{\Dk}$ the distance uncertainties\footnote{Note that the full distance uncertainties within each pixel may be non-Gaussian, but we assume Gaussianity here for simplicity.} for the galaxy and the pixel\footnote{see \citep{Singer2016_supplement} for a treatment of ansatz distances}\\

\noindent For each contained galaxy, the enclosing pixel's original 2D probability, $\ptDk$, is also used as a weight. These three weights are combined to produce the relative ranking for each galaxy, $\Wkg$, where $n$ is a normalizing factor. The product of $\Wkg$ and $\Pgals$ is the fraction of probability assigned to the $g$th galaxy within the $k$th pixel, $\Pkg$,

\begin{align}
    \Wkg &= n^{-1}\, \Ltildkg\, \ptDk\, \tilde{D}_{k, g}, \\
    \Pkg &= \Wkg\, \Pgals .
\end{align}

\noindent Finally, for each pixel, all contained galaxy probabilities are summed with $\pptDk$ to produce the final pixel probability, $\ppptDk$,

\begin{equation}
    \ppptDk = \pptDk + \sum_{g} \Pkg.
\end{equation}

This transformation redistributes the probability in the map proportionally to each pixel's completeness and \emph{concentrates} probability in regions that contain luminous galaxies within the GW-derived luminosity distance interval. This creates a new \teglon-processed version of the map that conserves the total probability of the original map,
\begin{equation} \label{eqn:conservation}
    \sum_{k}\ppptDk = \sum_{k}\ptDk = 1.0.
\end{equation}

\subsection{Pixel-Level Upper Limit Calculations} \label{sec:upper_limits}

\teglon\ also uses the HEALPix representation to estimate the detection efficiency of a model given a set of observations, {\it per pixel}. The footprint of each instrument in an observation manifest (Section~\ref{sec:implementation}) is modeled as a polygon, and we use the {\tt healpy} library to retrieve all map pixels within each footprint. To calculate the upper limit, we consider two types of quantities: those that are temporally invariant, and those that change with each observation. Section~\ref{sec:weighting} describes the invariant properties, $\ppptD$, $D$, $\sigma_{D}$, and $E(B-V)$. Mutable properties depend on each image in the observation manifest: filter, $\lambda$; MJD, $t$; and limiting magnitude, $m(t, \lambda)$.

Considering the footprint of all images of filter $\lambda$ at time $t$, there are $i$ enclosed pixels ($i \leq k$); for each we can consider an arbitrary model light curve, $M(t, \lambda)=M_{\lambda}(t)$, and calculate the maximum distance $D_{i, M, \lambda}(t)$\footnote{Note that we do not account for time-dilation or redshift effects at large distances, although these may contribute significant systematic uncertainty to our modeling near our maximum distance of 1.2~Gpc.  We leave updates to the algorithm for these effects to future work.} at which it could be detected in pixel $i$ given $m_{i}(t, \lambda)=m_{i, \lambda}(t)$,

\begin{align}
\mu_{i, M, \lambda}(t) &=  m_{i, \lambda}(t) - M_{\lambda}(t) - A_{i, \lambda}, \\
D_{i, M, \lambda}(t)~\mathrm{(Mpc)} &= 10^{0.2\, [\mu_{i, M, \lambda}(t) - 25]},
\end{align}

\noindent where $\mu_{i, M, \lambda}(t)$ is the distance modulus and $A_{i, \lambda}$ is the Milky Way\footnote{Note that we assume zero host extinction, which may not be true for certain GW counterparts.} extinction for a given sky position and filter. We integrate the pixel distance posterior up to $D_{i, M, \lambda}(t)$. Treating the line-of-sight distance marginal as approximately Gaussian in $D$ with mean $\Di$ and standard deviation $\sigma_{\Di}$, we write the contribution to the detection probability as
\begin{equation} \label{eqn:dist_integral}
\begin{split}
P_{i, M, \lambda}(t) &= \frac{n\,\ppptDi}{\sigma_{\Di}} \\
&\times \int_0^{D_{i, M, \lambda}(t)} \exp\Biggl[-\frac{1}{2} \left(\frac{D - \Di}{\sigma_{\Di}}\right)^{2}\Biggr] \,\mathrm{d}D ,
\end{split}
\end{equation}
where $P_{i, M, \lambda}(t)$ is pixel $i$'s contribution at epoch $t$ and $n$ is fixed so that the integral of the right-hand side from $D=0$ to $D=\infty$ equals $\ppptDi$ (i.e., $n$ absorbs the Gaussian normalization and any truncation at $D=0$).

To calculate the combined detection efficiency of $M_{\lambda}(t)$ across all $t$ for $i$, we take the complement of the product of the complements (i.e., the probability of detecting the model in at least \emph{one} epoch equals one minus the probability of nondetection in \emph{all} epochs). Summing over $i$ yields the total detection probability $P_{M,\lambda}$,

\begin{equation} \label{eqn:complement_product}
P_{M,\lambda}=\sum_{i} \ppptDi \left[ 1- \prod_{t}\left(1 - \frac{P_{i, M, \lambda}(t)}{\ppptDi}\right) \right].
\end{equation}

To combine the detection efficiency of multiple filters for a model, $M(\lambda)$, we sum the processed 2D probability for all pixels covered by any observation,

\begin{equation} \label{eqn:complement_product_total}
P_{{\rm obs}}=\sum_{i}\ppptDi,
\end{equation}

\noindent and repeat the process in Equation~\ref{eqn:complement_product},

\begin{equation} \label{eqn:model_prob}
P_{M}=P_{\rm obs}\left[ 1 - \prod_{\lambda}\left(1 - \frac{P_{M,\lambda}}{P_{\rm obs}}\right) \right].
\end{equation}

To the extent that multiple observations overlap with a given sky position $i$, we only decrement the probability once when a given observation would have detected the model in a particular filter and time.